\documentclass[aps,pre]{revtex4}
\usepackage{graphicx}
\begin{document}
\title{Effective macroion charge and stability of highly asymmetric 
electrolytes at various salt conditions}
\author{Vladimir Lobaskin\footnote{Present address: Max Planck Institute 
for Polymer Research, D-55128
Mainz, Germany. e-mail: Lobaskin@mpip-mainz.mpg.de}}
\affiliation{Physics Department, University of Fribourg,
CH-1700 Fribourg, Switzerland}
\author{Khawla Qamhieh}
\affiliation{Physics Department, College of Science and Technology,
Al-Quds University, Jerusalem, Palestine}
\date{\today}
\begin{abstract}
We  study  electrostatic   mechanisms  of  destabilization  of 
highly asymmetric electrolytes. For this  purpose, we perform
primitive model Monte Carlo  simulations of charged macroions 
immersed in multivalent salt solution. At  low  salt concentration,
the macroion effective charge  is reduced due to multivalent
counterion adsorption. At high salt  concentrations, the macroions 
become overcharged so  that their  apparent charge  has the opposite 
sign  to the  stoichiometric  one.  The inverted charge  is growing 
up to a saturation value  upon   further  increasing  the
salinity. The system remains stable at low as well as at very high
salt concentrations.  In the  intermediate  region,   close to the 
macroion  isoelectric  point,  we  observe macroion  aggregation. 
The obtained  phase  behaviour  closely resembles 
polyelectrolyte-induced instability of colloidal dispersions.
\end{abstract}
\maketitle

%\begin{galley}

\section{Introduction}
\label{sec:I}  

Electrostatic interactions in colloidal suspensions and their relation
to their  phase stability have  always been attracting  much attention
because  of  their  occurence  in many  biological  and  technological
circumstances. It has been known for  a long time that presence of the
oppositely  charged polyions seriously  deteriorates the  stability of
colloidal  dispersions.   In  polyelectrolyte-induced  aggregation,  a
correspondence between the location of the fast aggregation regime and
the   colloid  isoelectric  point   has  been   documented  repeatedly
\cite{Gregory,Miclavic,Grant,Borkovec}.        Furthermore,        the
electrophoretic  mobility of colloidal  particles, which  reflects the
apparent particle charge, has been shown to be tightly correlated with 
the 
system's stability.  The attractive interactions due to surface charge
heterogeneities have been thought to promote colloidal aggregation in
colloid-polyelectrolyte   systems  ("charge   patch   flocculation")
\cite{Gregory,Miclavic,Borkovec}.   It  however  only recently  became
clear that  apart from specific binding between  different solutes and
van der  Waals forces,  the electrostatic correlations  themselves can
lead to variety of  exciting phenomena in systems contaning oppositely
charged  polyions  such  as  giant  charge  inversion  or  short-range
attraction between  like-charged polyions (meaning  either colloids or
polyelectrolytes) \cite{Guldbrand,Gronbech,RB,Hribar,LowenAl,
Kjellander,Bratko,PRL,Shklovskii,Netz,Grosberg,Messina,Grosberg1,
Levin,VlachyRev,HansenRev}. Moreover,     computer
simulations and theoretical studies have revealed that the correlation
induced  attraction can  cause  colloidal aggregation  even when  just
small          multivalent          ions          are          present
\cite{Hribar,PRL,Shklovskii,Messina,JCP2}.   It  was  shown  that  the
universality  of the  underlying phenomena  is based  on  the enormous
energetic contribution of the Coulombic interactions to the solution's
free energy,  which dominates  over the entropy  and makes  the system
unsensitive   to   the   internal   structure   of   the   counterions
\cite{Kjellander,Shklovskii,Jure,Marie,Marie2,Carlsson}. A phase diagram
featuring a  phase separation  region around the  macroion isoelectric
point with  a consequent re-stabilization  was predicted for  a purely
Coulombic  system  containing  polyions  and a  sufficient  amount  of
multivalent counterions \cite{Grosberg,Toan}.

In the present work, we  present a computer simulation study of charge
inversion and phase instabilities in an asymmetric electrolyte treated
by a multivalent salt. This model is meant as a generic representation
of a solution containing two  types of oppositely charged polyions. We
hope to recover the basic properties of such a system originating from
the  electrostatics  and  translational  degrees  of  freedom  of  the
solutes.   Our  simulation   setup  resembles  a  common  experimental
situation  when  an initially  stable  suspension  is destabilized  by
adding  oppositely charged multivalent  ions (or  polyelectrolyte) and
aimed  to outline  the range of  activity of  the  ``patch charge
attraction''  \cite{Gregory,Miclavic,Borkovec} as  well as  assess the
theoretical   predictions    \cite{Grosberg,Toan,Harnau}   for   phase
behaviour of  asymmetric Coulombic systems  at different electrostatic
coupling parameters.   This work also complements  the recent computer
simulation  studies   of  these  phenomena  
\cite{JCP2,Per1,Marie2,Carlsson}  by  the investigation into salt
effects and the role of the effective macroion charge.

The paper  is organized as  follows. In Section \ref{sec:II},  we give
the description of the model  and parameter settings for the numerical
experiment.  Section \ref{sec:III}  gives  a detailed  account of  the
results  for  three  series   of  simulations  with  different  charge
asymmetries.   Section  \ref{sec:IV}   presents   discussion  of   the
mechanisms of  observed phase behaviour and its  relation to effective
interactions between  the colloids. The  conclusions are given  in the
final section.

\section{Model and method}
\subsection{Model}
\label{sec:II}

The systems under  consideration are asymmetric electrolytes described
within the framework of the primitive model. The reference electrolyte
contains two  types of spherical  charged particles: (i)  macroions of
diameter $\sigma_{M}=40$~\AA{}  and charge  $Z_M =-60$ and  (ii) small
ions  of   diameter  $\sigma_{I}=4$~\AA{}  and  charge   $Z_I  =  +1$,
representing  the counterions,  whereas the  solvent enters  the model
only  by  its relative  dielectric  permittivity $\varepsilon_r$.  The
added  salt consists  of  small ions  of  diameter 4\AA{}:  monovalent
anions $Z_a = -1$ and  cations of different valencies. We classified the
simulated systems  by the added cation charge: the series  A contains
cations with  $Z_c = +1$, the  series B $Z_c  = +3$, and the  series C
$Z_c   =+5$.  The  two last chosen  coupling strengths belong  to the
instability  region  on  the  generic  phase diagram  for  the  system
including only  macroions with their  counterions \cite{PRL,Per1}. The
initial 60:1 system was  previously studied in \cite{Per1,JCP1} and in
deionized  system with  trivalent counterions  a phase  separation was
found \cite{PRL,JCP2}. We characterize the  amount of added salt by a
ratio  of the  overall added cation  charge to  the overall  macroion
charge, $\beta =  Z_c \rho_c/( Z_M \rho_M)$, where  $\rho_i$  is the
number density of the corresponding species.

The  interaction  between  the  particles  in our  model  is  pairwise
additive, and for  pair $ij$, where $i$ and  $j$ denote either polyion
or counterion, it is given by

\begin{equation}
\label{eq:coulomb}
U_{ij}(r)= \left\{ \begin{array}{cc}
\infty, & r<{(\sigma_i+\sigma_j)/2} \\
\frac{Z_iZ_je^2}{4\pi\epsilon_0\epsilon_r}\frac{1}{r},&r\ge{(\sigma_i+
\sigma_j)/2},
\end{array} \right .
\end{equation}
where  $r$  is the  center-to-center separation between the particles.

To characterize the intensity of the electrostatic correlations between
counterions on the surface, we  use  the  counterion-counterion 
coupling  parameter $\Gamma  =  Z_I^2 l_b/a_Z$, where $l_b = (e^2/4 \pi
\varepsilon_0 \varepsilon_r k T)$ is the Bjerrum length, $e$  is the
elementary charge, $\varepsilon_0$ the dielectric permittivity  of
vacuum, and  $a_Z= [Z_I/(\sigma/e)]^{1/2}$ the  average  distance 
between  two neighboring  counterions  at  the charged surface
characterized by the surface charge density $\sigma$. It is known that
the correlation-induced attraction appears at $\Gamma > \Gamma^* \approx
2$ \cite{RB,JCP2,Per1}. Our systems A, B, and C  correspond  to  
$\Gamma=0.8$,  4.1, and  8.8, respectively. 

The systems  are considered at a  fixed macroion number  density
$\rho_M=2.5 \cdot  10^{-7}~$\AA{}$^{-3}$ corresponding to a macroion 
volume fraction $\phi_M=0.0084$. A temperature of  $T=298$ K and  a
relative dielectric  permittivity of $\varepsilon_r=78.4$  were used.
For each of the  three  series, 11 values of $\beta$ ranging  from  0.1
to 10 were studied. 

\subsection{Method and simulation settings}

In this study, we applied multiparticle Monte Carlo (MC) simulation
method, which was developed in our earlier works
\cite{JCP1,Springer}.  The  canonical  NVT  ensemble  with   periodic 
boundary conditions and  the  Ewald   summation  for  handling  the 
electrostatic interactions were employed. The simulated bulk systems
were composed of 20 macroions, 1200 monovalent counterions, and
various  amounts of added salt ions. Typically, $0.5-1 \cdot 10^5$
attempted MC moves per particles were made in the production runs,
while for samples at $\beta \approx 1$ we performed up to $1 \cdot
10^7$ moves.

For the simulation of solutions with much salt we combined cluster MC
moves  for macroions  \cite{JCP1} with  {\em swap  moves}, which  is a
modified implementation  of the  cluster moves. In  the swap  move, we
first  select  the  macroion  and  determine the  particles  within  a
concentric sphere around  it. Then a second sphere  of the same radius
is  taken   centered  at  the  destination  vector   of  the  macroion
translational move $\delta  {\bf r}_M$ and a central  inversion of the
content  of the  whole volume  within the  two  (possibly overlapping)
spheres  about the  center of  symmetry at  $\delta {\bf  r}_M  /2$ is
performed. The acceptance ratio is  calculated then in the same way as
for  the  standard  cluster  move  \cite{JCP1}, which  results  for  a
hard-sphere system and with 100\% probability of the particle inclusion
in the cluster in rejection of the moves that change the number of
particles in  the cluster (cf. discussion in \cite{JCP1}). This  type 
of move significantly improves
the acceptance  ratio for  the systems with  homogeneously distributed
ions in the bulk as the macroions always move into an ion-free volume.

To evaluate the effective macroion charge and the osmotic coefficient,
we applied the  spherical Wigner-Seitz (WS) cell model.  The model was
solved by  canonical MC simulations for one  centrally placed macroion
with  corresponding amount  of small  ions, while the other settings
were the  same as for the bulk  simulation. The simulation  package
MOLSIM v3.1 by Per  Linse et al. was used throughout. Further details
of the simulation protocol can be found in Refs. \cite{JCP1,Springer}.

\section{Results}
\label{sec:III} 
\subsection{Single-macroion properties. Thermodynamics}
 
\begin{figure}[htb]
\begin{center}
\vskip 0.1in
\includegraphics[height=6cm,width=8cm]{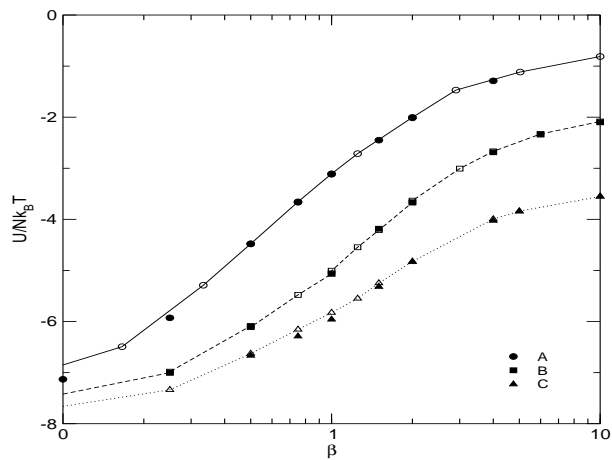}
\end{center}
\caption{Reduced electrostatic energy for series A, B, and C from bulk MC
simulation with 20 macroions (full symbols) and the corresponding WS
cell model (open symbols).}
\label{fig:1a}
\end{figure}

\begin{figure}[htb]
\begin{center}
\vskip 0.35in
\includegraphics[height=6cm,width=8cm]{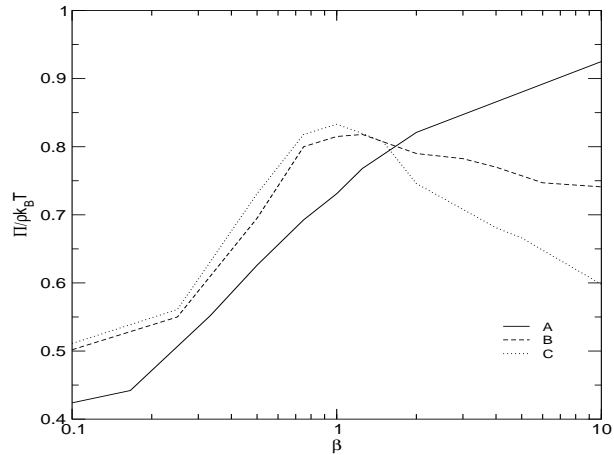}
\end{center}
\caption{Osmotic coefficient ($\rho(R_{cell}/\rho)$) for  series A, 
B, and C from MC simulation of the WS cell model.}
\label{fig:1b}
\end{figure}

We   start  discussing  the   numerical  results   from  thermodynamic
properties  of  the  chosen  electrolytes  as  function  of  the  salt
concentration.   In  Figs.~\ref{fig:1a}-\ref{fig:1b}  we  plotted  the
reduced electrostatic energy of the  solution $U/N k_B T$, and the
osmotic coefficient  $\Pi /\rho k_B  T$, where  $N$ and  $\rho$ are 
the total number  and the  number density  of the  ionic  species,
respectively, $k_B$ the  Boltzmann constant, $\Pi$ the osmotic
pressure. The reduced energy  was calculated both from bulk 
simulation and WS cell model. The Fig.~\ref{fig:1a} shows that  the 
total potential  energy  is  negative  at all  the  studied conditions
as a result of  strong attraction between the macroions and
counterions.  Its  magnitude is decreasing monotonously  with the salt
content,  which is  a  simple consequence  of  increasing the  overall
number of  particles. All the  curves have similar shape  with varying
curvature.  An inflection point is observed at about $\beta= 1$ on
each curve. For the osmotic  coefficient shown in Fig.~\ref{fig:1b} we
present the cell model results obtained from the total density of ions
at the  cell boundary $\Pi_{cell}/(\rho k_B  T)=\rho (R_{cell})/\rho$.
The osmotic coefficient behaves  differently for systems A,  and B
and  C. For the system A,  it is increasing roughly proportionally  to
$\ln{\beta}$ up to $\beta =  1$, and after that is more  slowly
approaching unity, the ideal solution  value.  In the two  latter
series it  is again growing proportional  to $\ln{\beta}$  at low 
salt but  with  slightly higher coefficient. Both curves for series B
and C have a peak at $\beta = 1$ and then decrease proportionally to
$-\ln{\beta}$, the decay of series C results  being faster. The value 
of the osmotic  coefficient at the peak is about 0.8. The proximity of
the $\beta$ values to unity in all these  cases  reflects the  high 
fraction of  the  free  ions in  the solution. This fraction can be
roughly estimated as $(N_a+N_c+Z_{eff}+N_M)/N$,  which  gives  at 
$\beta=1$  for  complete adsorption  of the  added cations  121/181,
121/141,  and  121/133, or 0.67, 0.86, and 0.91 for the systems A, B,
and C, respectively. The further decrease of the osmotic coefficient
is a result of association  of salt anions with the multivalent
cations, which reduces the number  of "free" species as compared to
the ideal solution value.

\subsection{Single-macroion properties: Effective charge}
 
The concept  of effective polyion  charge attracted much  attention in
literature  due  to  its  importance  for the  reduction  problem.  It
immediately arises whenever  the solvent- or ion-averaged interactions
between  solutes are concerned.  The two  major trends  (excluding the
experimental  definitions)  for  two-component systems  (macroion  and
counterions)   are   represented  by   the   thermal  and   structural
definitions. The different definitions in a spherically symmetric case
regard the distance from the  macroion center, within which the charge
should  be  counted.  The  structural  definitions are  based  on  the
knowledge of  the radius,  at which the  mixed macroion-ion  rdf falls
below  unity,   or  other  criteria.  The  simplest   of  the  thermal
definitions counts  the charge inside the  macroion-ion binding radius
($|U(r_{therm})| = kT$). The more accurate one considers the chemical
potential of the counterion in  the macroion field. In the cylindrical
case  it  was  developed  by  Manning and  Oosawa 
\cite{Manning,Oosawa} and adopted for spherical geometry by Belloni
\cite{Belloni}. The radius in this approach is measured at  the 
inflection point  of the accumulated running charge  curve plotted over
inverse distance  from the macroion center  $1/r$ (see also
refs.\cite{Belloni,Deserno} for discussion of alternative definitions).
Obviously, the radius  is  different for the  ions of different
valencies.

\begin{figure}[htb]
\begin{center}
\includegraphics[height=18cm,width=8cm]{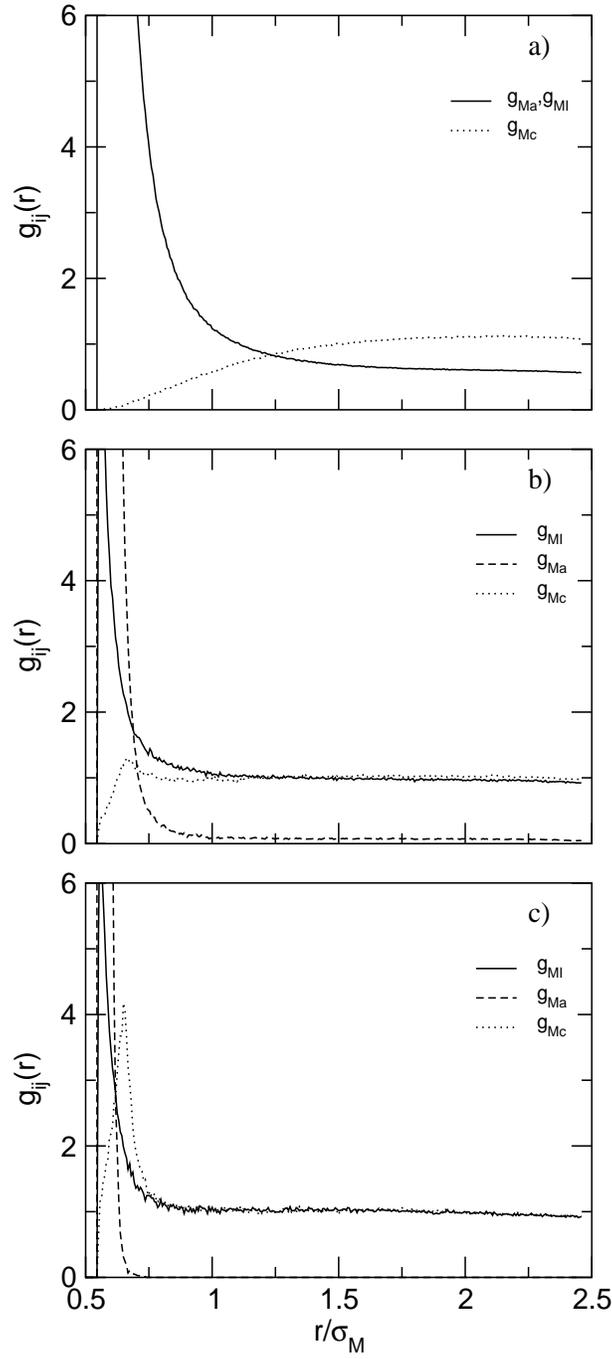}
\end{center}
\caption{Counterion ($g_{MI}$) and added salt ion distribution functions
(multivalent counterion $g_{Mc}$, coion $g_{Ma}$)  for the series A, B,
and C, respectively, as obtained from MC solution of the WS cell model.}
\label{fig:2}
\end{figure}

In the case  of multicomponent electrolyte, however, we  have to apply
different definitions to different  ionic species due to qualitatively
dissimilar behaviour of the integrated running charge depending on the
salt valency.  Here,  we use MC simulation to solve  the WS cell model
\cite{JCP1}   and  then  extract  the  effective  charge  out  of  the
integrated running charge curves. For series  A and series B, C at low
salt,  we apply the  inflection point  method \cite{Belloni}.  For the
At $\beta>1$, where the inflection point was  not accessible
anymore, we  evaluated the effective  charge at the first  extremum of
the integrated running charge curves.

The Fig.~\ref{fig:2} demonstrates  typical ionic distribution profiles
for systems A, B and C. While all the curves decay monotonously in the
series  A, the  systems  B and  C  display diverse  behaviour for  the
different ionic species. The  number density of monovalent counterions
decays similarly  to the system A.  The multivalent ion  density has a
high maximum at  the polyion surface, while farther in  the bulk it is
close to zero.  The coion distribution is peaked  close to the maximum
of the multivalent ion distribution because of their attraction to the
macroions with the inverted charge  and becomes very pronounced at the
high salt concentrations \cite{Tanaka}.
\begin{figure}[htb]
\begin{center}
%\leavevmode
\includegraphics[height=18cm,width=8cm]{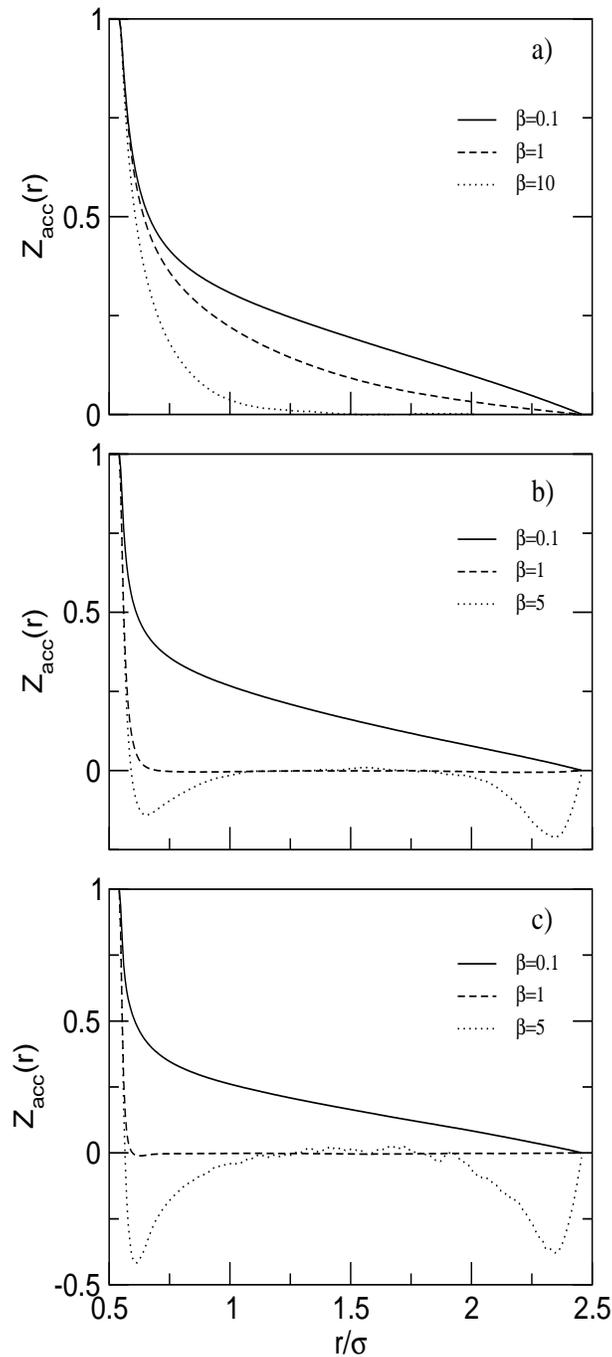}
\end{center}
\caption{Accumulated running charge as a function of the distance from
the macroion center as obtained from MC solution of the WS cell model,
a) series A, B series B, c) series C.}
\label{fig:3}
\end{figure}

The    integrated   running   charge    curves   are    presented   in
Fig.~\ref{fig:3}. There  is a distinct  qualitative difference between
system A  and two other  systems B, C.  While in the former one the total
charge  is decaying  monotonously and  reaches zero  only at  the cell
boundary, in the two latter systems the charge exhibit a sharp drop to
zero within  a very narrow  region $r<1.2\sigma_M$. At  $\beta=1$, the
running charge remains  close to zero up to the  cell boundary. At the
higher $\beta$, the charge is changing its sign and after that decaying
slowly.  Close to the  cell boundary,  one can  see the  cation charge
build-up  (note that  the charge  is divided  by the  negative $Z_M$),
which originates from repulsion  between the charge inverted macroions
and cations.   Note that  the accumulated charge  in series A  at high
salt concentrations reaches zero already  at $r = 3\sigma_M$, which is
much less than the mean distance between the macroions. 

\begin{figure}[htb]
\begin{center}
\vskip 0.1in
\includegraphics[height=6cm,width=8cm]{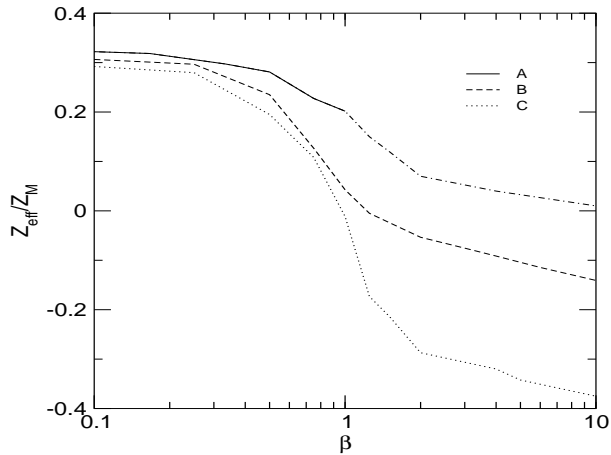}
\end{center}
\caption{Effective macroion charge for the series A, B, and C  vs added
salt concentration $\beta = Z_c \rho_c/(Z_M \rho_M)$ as obtained from MC
solution of the WS cell model. The data for system A at $\beta>1$
(dotted-dashed curve) are obtained in the region, where the definition of
the effective charge based on the inflection point criterion is no longer
justified.}
\label{fig:4}
\end{figure}

In  Fig.~\ref{fig:4}, we present  the effective  macroion charge  as a
function  of  $\beta $  for  all three  series.  For  all the  systems
$Z_M^{eff}/Z_M$ starts from about 0.3 at low added salt (its value for
the   60:1    electrolyte   \cite{JCP1})    and    then   monotonously
decreases. The effective  charge changes its sign for  systems B and C
and regains roughly the same magnitude  (0.17 for series B and 0.36 for
series  C)  on  the negative  side.  It  is  interesting to  note  the
qualitative similarity of  all the three curves. All  of them have two
relatively flat regions at very  low and very high salt concentrations
and a region of fast variation with an inflexion point at about $\beta = 
1$. The  curve for series  A asymptotically  approaches but  never
crosses the zero line. Very slow dependence of the effective charge on
the salt content at $\beta > 1$ follows from logarithmical increase of
the counterion entropy with counterion number density \cite{Toan}. One
should note that the decrease of the effective charge with the salt 
concentration, as it is seen for $\beta<1$, is characteristic for dilute 
and deionized systems, where the effective charge is far from its 
saturated value \cite{Belloni}. At the higher monovalent salt content, 
$\beta>1$, the ionic double layer saturates and  $Z_M^{eff}/Z_M$ starts
growing with salt concentration towards unity  \cite{Belloni,Trizac}. We
could not see this upturn due to the failure of the inflection point
criterion in this region.  The failure is due to the fact that the
underlying  division of the ionic population into the "condensed" and
"free" parts is no longer valid. The inflection point can still be
detected formally but it refers now to the far field properties of the
double layer (the inflection point is located next to the cell
boundary) and does not reflect its short-range decay, which is
important for the interaction between the macroions.

\subsection{Collective macroion properties: Structure}

The   object   of  our   main   interest   here   is  the   macroionic
distribution. We therefore analyse only macroion-macroion  partial
distributions and  extract  those  data   that  characterize  the 
macroion  density fluctuations. Namely,  we consider the  long-wave limit
value  of the macroion-macroion  partial structure  factor $S_{MM}(q)$, 
which would quantify  the  reduced  osmotic  compressibility  in  a 
corresponding one-component  fluid of  macroions.  We  should note  that
due  to the small  size of  the simulated  sample  we can  access only 
relatively short-wave region  ($q\sigma_M > 0.01$) and therefore  have
only rough estimate of  $S_{MM}(0)$. One should  expect stronger
deviations from unity for the true  $S_{MM}(0)$, i.e. even lower
values for the stable solutions and the higher ones for the aggregating
systems.

\begin{figure}[htb]
\begin{center}
%\leavevmode
\includegraphics[height=7cm,width=8cm]{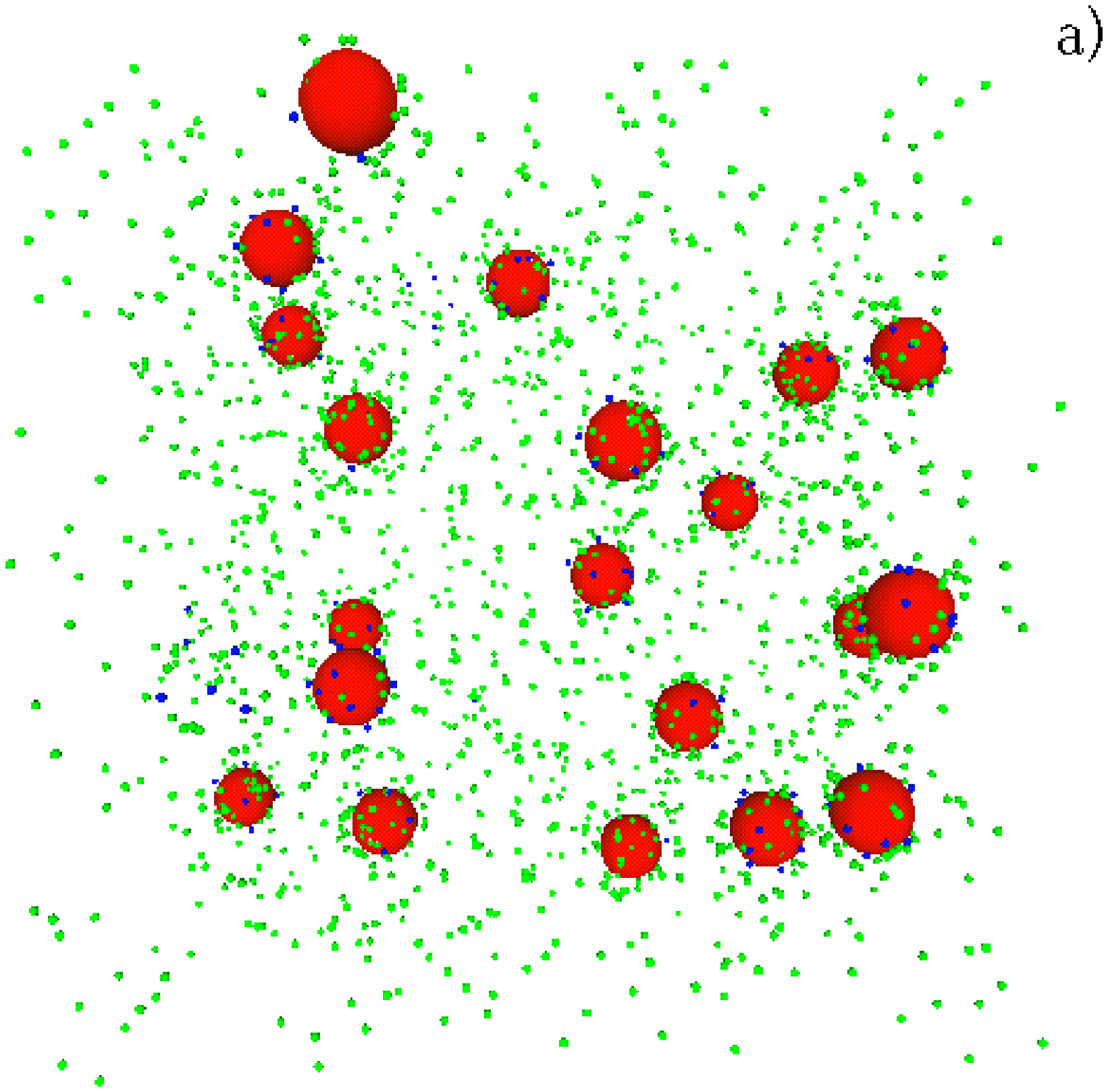}
\includegraphics[height=7cm,width=8cm]{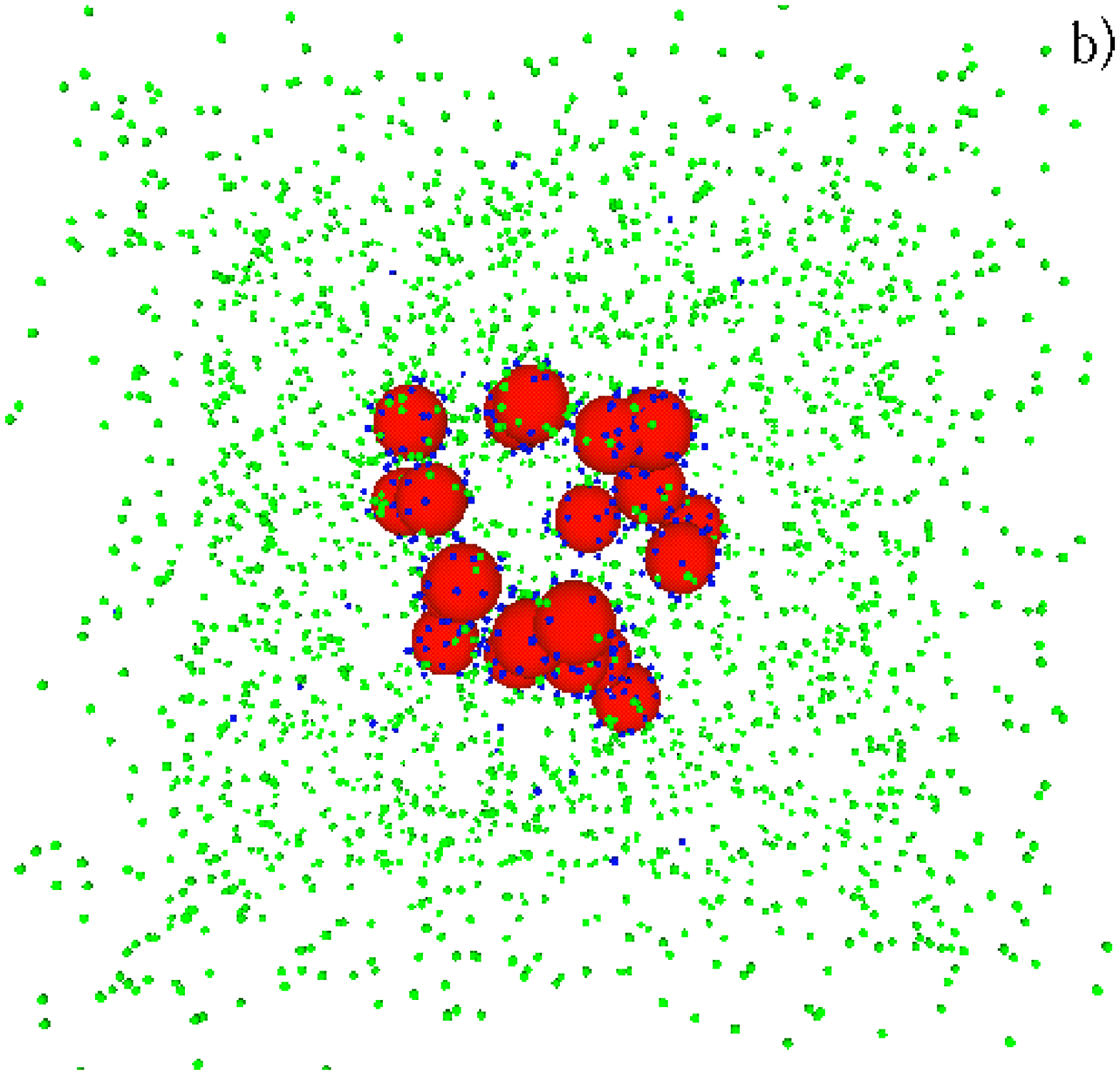}
\includegraphics[height=7cm,width=8cm]{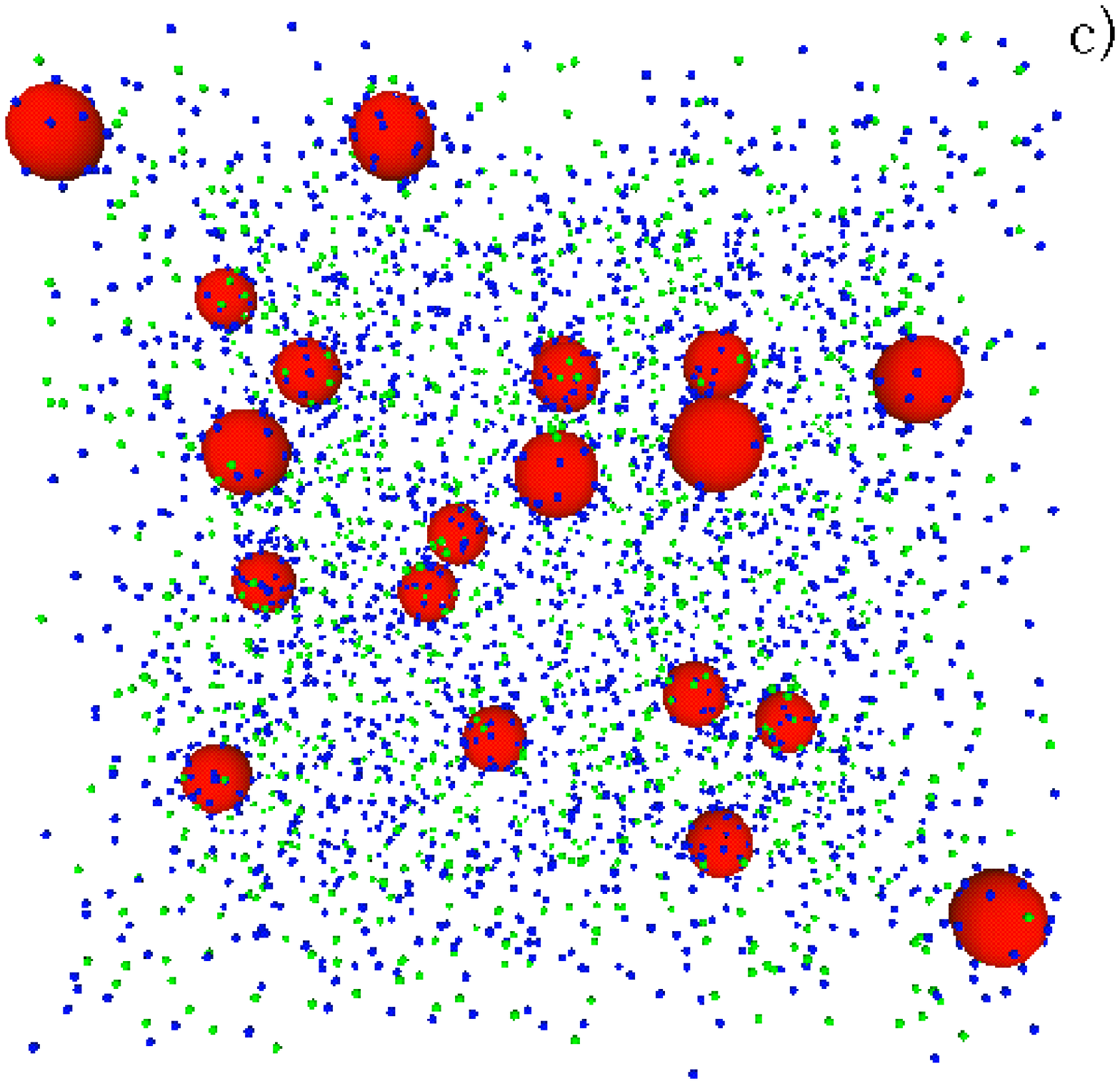}
\end{center}
\caption{Snapshots of the primary simulation box for the series C with
20 macroions taken at the end of simulation at a) $\beta = 0.1$, b)
$\beta = 1$, and c) $\beta = 5$. The big spheres are representing the
macroions, the small ones representing the counterions and added salt 
ions. The coions are omitted in c) for clarity. In the snapshot b) 
corresponding to $\beta = 1$, one can see macroion cluster.}
\label{fig:5}
\end{figure}

The  macroion  distribution  in  the  bulk  solution  can  be  easiest
illustrated  by simulation  snapshots shown  in  Fig.~\ref{fig:5}. The
snapshot  a) corresponds to  a low  added salt  content $\beta=0.1$
and  reveals a homogeneous spatial distribution of  the macroions. In
the snapshot b) corresponding  to $\beta=1$  one can  see macroion 
clusters.  All the macroions are in aggregated  state.  Finally,
snapshot c) demonstrates that at high salt $\beta = 5$ most macroions
are free and the solution is  back  to  the stable  state.  The
cluster re-dissolution is in  agreement with theoretical predictions 
\cite{Toan} and recent simulations for polyelectrolyte-induced  a
aggregation of colloids and globular proteins \cite{Marie2,Carlsson}. 
We should note that even at high $\beta$  values occasional particle 
pairs are  formed. To  ensure the equilibrium thermodynamic  state of
the system  we performed additional simulations with  initial compact
configuration of  the macroions. The macroions were  allowed to  move
after the  equilibration of  the ionic subsystem.  Then,  indeed  we 
observed stable  aggregates  at  $\beta \approx  1$ and 
re-dissolution of  the macroions  at the  higher salt dose. The
parameters of the aggregates  were drifting slowly towards their
values from the simulation with  random initial configuration with an
increase of simulation length.  The problem with matching the two sets
of data for the  re-dissolution and the cluster-cluster aggregation is
caused by  the wide energy barriers originating from the large net
charge of the  clusters \cite{Toan2}. Although the cluster charge is
not growing  proportionally to the number of macroions due to the
partial counterion release, the cluster charge yet reached values of
$2 Z_M^{eff}$ in our systems B and C at $\beta>1$.

\begin{figure}[htb]
\begin{center}
\vskip 0.1in
\includegraphics[height=18cm,width=8cm]{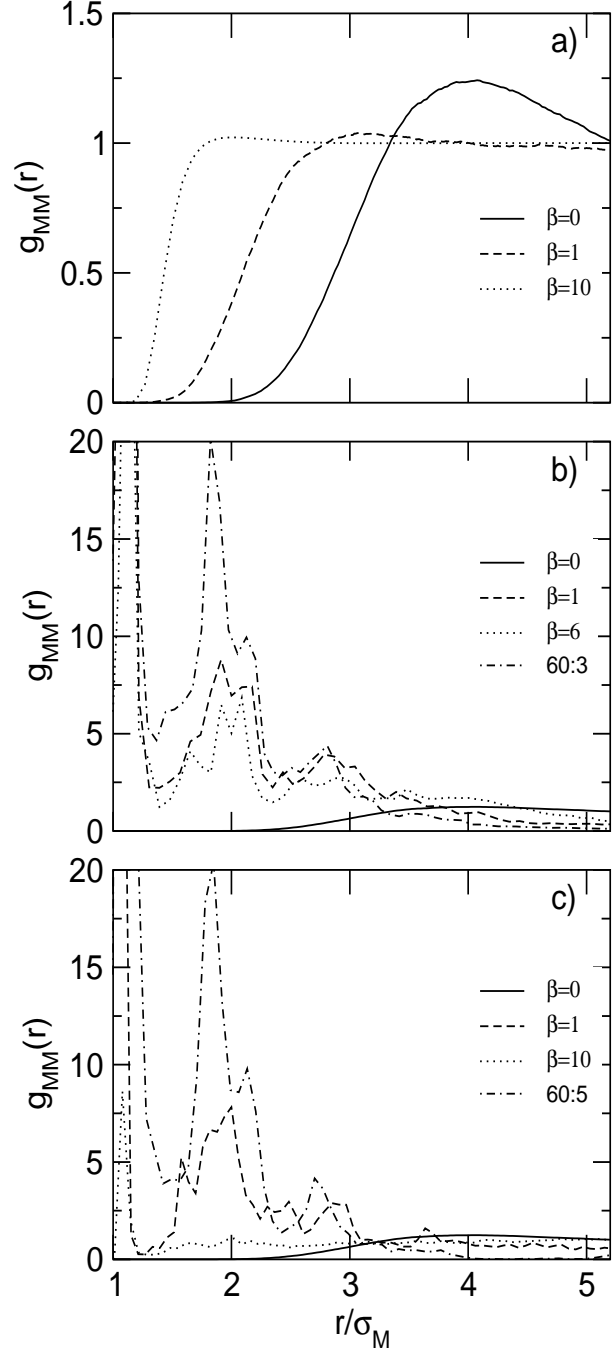}
\end{center}
\caption{Macroion-macroion partial radial distribution functions  for 
series A, B, and C, respectively, at the indicated added salt
concentrations given in units of $\beta = Z_c \rho_c/(Z_M \rho_M)$ from
MC simulation with 20 macroions. The curves obtained for asymmetric
electrolytes  60:3 and 60:5 without monovalent ions are given for
comparison (see discussion in section \ref{sec:IV}). The stripped initial
maxima have values of 60 and 150  for the series B and C, and 200 and 230
for 60:3 and 60:5 systems,  respectively. }
\label{fig:6}
\end{figure}
\begin{figure}[htb]
\begin{center}
\vskip 0.2in
\includegraphics[height=18cm,width=8cm]{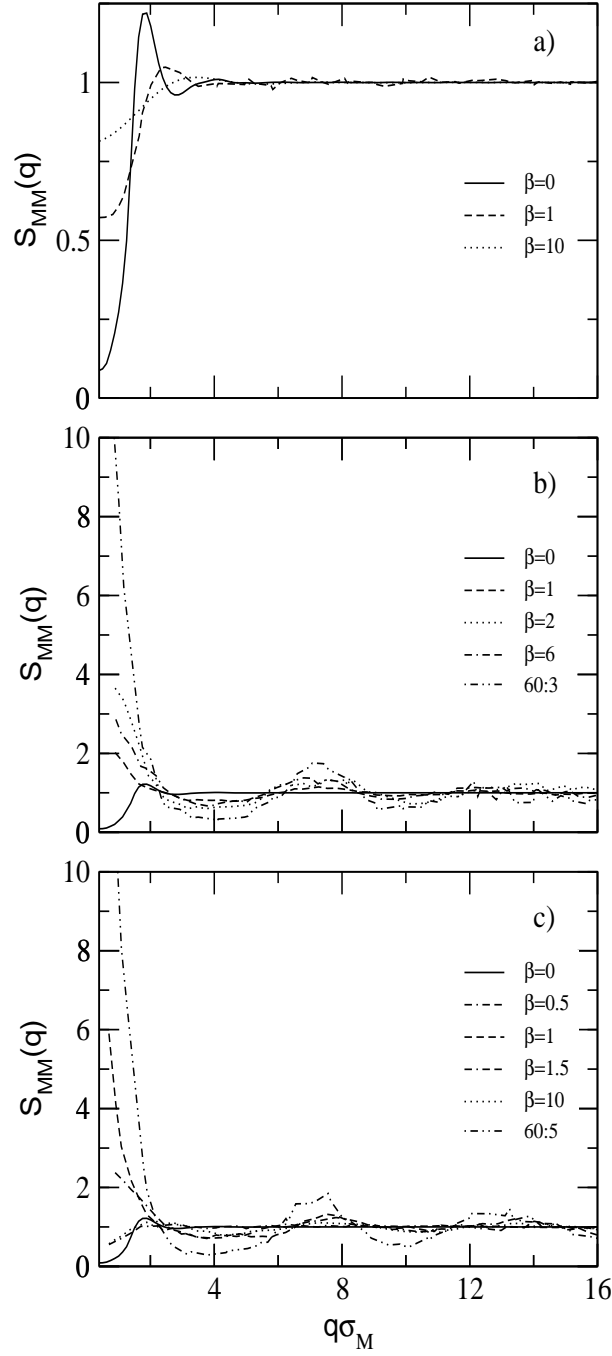}
\end{center}
\caption{Macroion-macroion partial structure factors for series A, B, 
and C, respectively, at the indicated added salt concentrations given in
units  of $\beta = Z_c \rho_c/(Z_M \rho_M)$ from MC simulation with 20
macroions.  The curves obtained for asymmetric electrolytes  60:3 and
60:5 without monovalent ions are given for comparison (see discussion in
section \ref{sec:IV}).}
\label{fig:7}
\end{figure}

Figure~\ref{fig:6}    presents   partial    macroion-macroion   radial
distribution functions  (rdf) for the three systems  at different salt
concentrations.  In Fig.~\ref{fig:6}a,  one  can see  the results  for
system  A,  which show  gradual  decay  of  the macroion  ordering  on
increasing   salt.  The  peak   of  the   rdf  appears   initially  at
$r=\rho^{-1/3}$ and  has a value  of 1.15, then moves  towards shorter
separations, while  decreasing at the same time  and almost disappears
at ($\beta > 1$). The  trend is readily understood from decreasing the
effective  macroion charge and  growing the  screening ability  of the
electrolyte at the higher salt  contents. For the remaining two series
B and C, the trend is very different. Already at $\beta=0.5$, there is
a  peak near  the  macroion  contact distance (or,  more exactly,  at
$r=\sigma_M+\sigma_c$), which  then grows high upon  salt addition and
reaches value of 60  at $\beta = 1$ for the series  B, and 150 for the
series C.   At $\beta  >1$ it considerably  drops in height  but still
remains  quite   high  even  at   $\beta=10$.  A  secondary   peak  at
$r=2\sigma_M$ corresponding to the  second particle layer in a cluster
is seen for some of the curves in  series B and C at  $\beta>1$. We would
like to note the reappearance of the remote peak at $r\sigma_M =5$ for
the series C at $\beta=10$. We made also another interesting observation
for the system C. At the salt concentrations $\beta \geq 1$, the main 
peak of the macroion-macroion rdf shifts towards $r=\sigma_M$ and is 
seen now at $r=1.02 \sigma_M$ instead of $r=\sigma_M+\sigma_c=1.1$. 
This effect might be related to strengthening the electrostatic depletion
contribution to the macroion interaction \cite{LowenAl}.

We  plotted  some  macroion-macroion  partial  structure  factors  for
different  salt concentrations  in  Figure~\ref{fig:7}. The  structure
factors  for  the series  A  again shows  decay  of  the structure  on
increasing  $\beta$.  For  the  two remaining  series,  the  structure
factors have initially  a peak at $q \sigma_M = 2  $ that corresponds
to the inverse mean distance between the macroions and a low value in
the long-wave limit. On increasing salt,  this peak disappears and two
new peaks develop instead at $q \sigma_M \rightarrow 0$ and $q\sigma_M
\approx 7$.  The first  peak reflects the  average macroionic  cluster
size  that grows with  the salt  addition  until  $\beta \approx 1$ 
for  series B  and C.  After  that, when  even  more  salt is  added, 
the  value of  the long-wave peak  decreases, which  indicates the
re-dissolution  of the clusters. The broad secondary  peak at
$q\sigma_M \approx 7$ corresponds to the macroion-macroion density
build-up close to contact distance.

\begin{figure}[htb]
\begin{center}
\vskip 0.2in
\includegraphics[height=6cm,width=8cm]{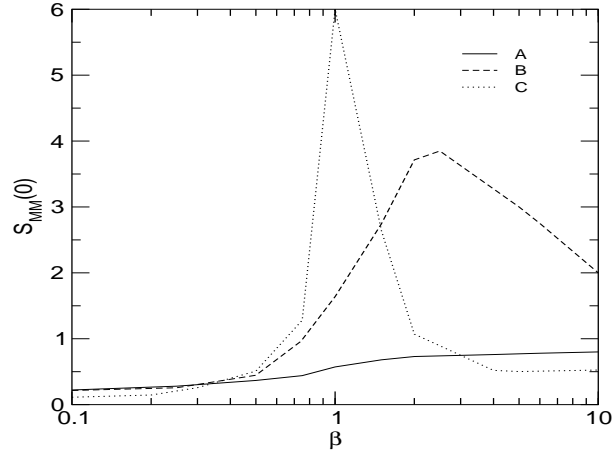}
\end{center}
\caption{Long-wave value of the macroion-macroion partial structure 
factor ($S_{MM}(0)$) for the series A, B, and C from MC simulation 
with 20 macroions.}
\label{fig:8}
\end{figure}

The whole set of values of the long-wave structure factor value $S(0)$
for  all three  series is  shown in  Fig.~\ref{fig:8}.  The  curve for
system A  shows gradual increase  between $\beta =0$  and 6 while the 
curves for  systems B  and C  have well  expressed  maxima around
$\beta=1$.  For  the system  B, the peak  is slightly  shifted towards
higher  salt  concentration  due  to  the   competition  between  the
adsorption energy  and the cation  entropy, which leads to incomplete
macroion charge compensation at $\beta=1$. For  the series C, the
$S(0)$ value returns to the  values lower than 1,  as it was  at
$\beta \ll 1$.  The overall trend for  the system C  corresponds to
its destabilization  at $\beta \approx 1$ a consequent
re-stabilization  at high salt that occurs due to recovery of the
repulsion between the macroions (cf.  the effective charge plot in
Fig.~\ref{fig:4}) \cite{Grosberg,Toan}.

\section{Discussion}
\label{sec:IV} 

As  we have  seen  from the  above  results, the  salt conditions  are
crucial  for the  stability of  asymmetric electrolytes.  On  the path
along  the  salt  concentration  axis,  two  important  parameters  of
macroion interactions are affected: the Debye screening length and the
effective charge of the macroions.  The effect of the former is rather
smooth and can  be detected through a simple  damping  of the
interactions upon salt  addition. The sensitivity of the  asymmetric
electrolyte to the added salt depends also  on the relative amounts of
the ``native'' and ``added''  ions. We  note that overall  properties
of  the studied systems  mark the  charge compensation  point $\beta=1$ 
either  by an extremum (the osmotic coefficient)  or an inflection
(potential energy, effective charge),  which can be  understood as a
crossover  from the counterion  dominated ($\beta <  1$) to  the salt 
dominated screening $\beta > 1$.

\begin{figure}[htb]
\begin{center}
\vskip 0.2in
\includegraphics[height=6cm,width=8cm]{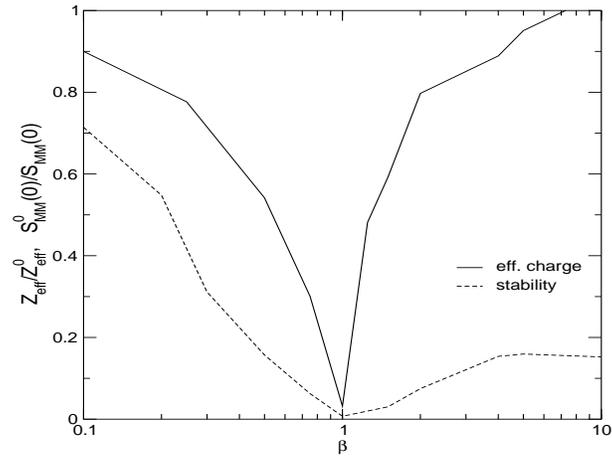}
\end{center}
\caption{Absolute value of effective macroion charge and the stability
ratio ($S_{MM}(0)^{-1}$) for the series C normalized by their values at 
$\beta =0$.}
\label{fig:9}
\end{figure}

Another interaction  parameter, the  effective  charge, influences
mostly  the   macroion  related   properties  such  as   the  macroion
distribution  and the  stability  of the  macroion  subsystem. In  the
counterion dominated  regime, the effective charge  is changing almost
linearly with the salt concentration.  This makes the system extremely
sensitive  to  the  addition  of  salt. In  contrast,  at  the  higher
electrolyte    strength   the    effective   charge    is   increasing
logarithmically with the amount of  added multivalent salt.  We note a
close  similarity  between  the  observed  charge  behaviour  and  the
macroion electrophoretic mobility vs added polyelectrolyte dose curves
reported  in Refs.\cite{Grant,Borkovec}. Fig.~\ref{fig:9}
demonstrates  the direct relation  between  the  effective  charge 
and the  stability  of  the electrolyte solution. In the plot  one can
see the relative effective charge   curve   together  with   the  
characteristic  of   stability $S_{MM}(0)^{-1}$  for  the series  C. 
It  is  important to  note  the approximate symmetry of the curves in
Fig.~\ref{fig:9} relative to the isoelectric  point, which  allows
one  to  formulate the  result in  a symmetric manner.  While one can
say  that on the  right-hand side of  the diagram the macroions are 
overcharged by  the excess multivalent  counterions, on its left-hand 
side there is  excess of macroions themselves  and they are 
overcharging the  multivalent  ions.  The  latter  gets a  richer
physical meaning when we  imagine polyelectrolyte molecules instead of
the multivalent ions. The  effect of the internal counterion structure
was discussed recently \cite{Jure,Marie2}.  As it was shown previously
for the asymmetric electrolytes without added salt \cite{Per1}, the
parameters  of importance are the charge asymmetry  between the
oppositely charged ionic species and  the ion correlation  parameter
$\Gamma$.   Extending  our conclusions, we  should expect that the
strongly  charged systems with similar  energetic parameters have 
common features  irrespectively of the   internal   structure  of  
the   polyions,   whether  they   are polyelectrolytes or colloids.

While changing  the salt  valency leads to  a smooth variation  of the
system's properties such as  the internal energy, osmotic coefficient,
or  effective macroion charge at the salt-overdosed conditions far
from  the isoelectric point,  the interactions and  structure change
drastically in all the series at $\beta =1$. Our results confirm that
the instability appears at a certain  threshold counterion valency and
becomes sharper at  the higher  counterion charge  (or counterion 
coupling $\Gamma$). Therefore, at the higher $\Gamma$ one can expect
on  one hand a much  stronger macroion aggregation at $\beta  =  1$ 
and,  on   another  hand,  a  much better expressed overcharging and 
re-dissolution effects.  Thus, the  three systems of our choice
correspond  to (i) the weak coupling  region and counterion entropy
dominated   behaviour  (series   A),   (ii)  border   region
demonstrating  the coexistence of condensed  and free counterions at
$\beta=1$ (series B) and (iii) the high coupling (or ``low 
temperature'') regime with completely condensed multivalent 
counterions (series C). The alllter case was  analyzed in  detail 
recently \cite{Toan}.   At the same time, the  instability in real
solutions should  be much enhanced due to van der Waals  forces, which
due to their short-range character similarly play a role only in the
vicinity  of the isoelectric point, i.e.  the whole sequence must be
shifted towards lower $\Gamma$ values \cite{Carlsson}.

Finally, we  would like  to report  on the simple  salt effect  on the
effective charge and the counterion correlation attraction. The salt 
effects in similar systems  were discussed  previously \cite{JCP2}.
It  was confirmed that the added  salt screens the repulsions between 
the macroions and leads  to  weakening  the  long-range  macroion
ordering.  In  a  60:2 electrolyte,  addition of  divalent  ions
caused  an  increase of  the electrostatic attraction. In the  60:3
electrolyte, which is similar to  our system B but has no monovalent
counterions,  no significant   effect was  noticed. Basing  on the
above  analysis,   we  can  explain  the  increase   of  the  macroion
aggregation in the 60:2 electrolyte by that the added divalent ions in
fact  lessen  the  effective  macroion  charge,  which  is  not
compensated  completely at $\beta =1$ (similar to system B of the
present work at $\beta=1$)  \cite{PRE}. The apparent salt  effect in 
the  60:3  electrolyte  was   small  because  the considered salt 
concentrations placed the system  in the center of the instability
region,  where  it  is  not  very  sensitive to  the  salt. The  salt
concentration range  in \cite{JCP2} corresponds to  $1<\beta<4$ in the
current work (cf. the curve for the system B in Fig. \ref{fig:8}) Now,
we repeat the comparisons of  Ref. \cite{JCP2} between the solutions B
and C  at $\beta =1$  and corresponding systems with multivalent ions
only, namely 60:3 and 60:5 electrolytes. 

In the 60:3 system, the absolute potential energy per macroion is
about  2.5\% smaller than in series B, while the difference is about
3.3\%   between the 60:5 electrolyte and the system C. Similar
relations hold  for the  partial electrostatic energies: the
macroion-cation and cation-cation energies, respectively, are 3.3\%
and 5.5\% higher in the 60:3 solution than in the B series at $\beta
=1$. For the 60:5 solution and system C, these diffrences are
marginal: the macroion-cation energy is 0.3\% greater in the former,
while the cation-cation energy is the same within the statistical
uncertainty.  The osmotic coefficient calculated in  the cell model in
the 60:5 system is equal to zero, which means that  all the 
multivalent  ions  are adsorbed  and  the bulk  osmotic coefficient is
then just $\Pi /(\rho_M  k_B T) \approx 0.08$. The simulation for the
60:3 electrolyte showed about 5 times higher cation concentration in
the bulk for the corresponding system B, though both concentrations
were  very small. At the same  time, the presence of the monovalent
ions leads to a very modest decrease in correlation between the
multivalent ions. The cation-cation correlation peak in system B
falls  by about 10\% with respect to 60:3 electrolyte, which
corresponds to a slight  decrease of the effective correlation
parameter  $\Gamma$. In the system C compared to the 60:5 system, the
analogous  decrease  is  just  about 5\%. The macroion-cation  rdfs
display marginal sensitivity to the added electrolyte. The peak height
for the salt-free case is only 5\% higher in the 60:3 system,  and 2\%
higher in the 60:5 system than for the corresponding samples B and C.

The  relative  effective charge  $Z_M^{eff}/Z_M$ for the both charge
asymmetries  increases in magnitude with the salt addition. With
trivalent ions, it rises  from $0.015 Z_M$ in 60:3 system to $0.03
Z_M$ in system B. For pentavalent ions, it changes from 0 in 60:5
electrolyte to  $-0.01 Z_M$ in system C. Note that the presence of 
the anion layer around the  adsorbed cations does not cancel the 
effect (see Fig. \ref{fig:2}).  A significant role here is played  by
the monovalent counterions, which  also accumulate close to the
macroion surface.  The macroion structure displays considerable
sensitivity  to  the presence of monovalent ions. In  the 60:3 system,
the  first  peak of  the macroion-macroion  rdf increases from 60 to
200 as compared to the system B (Fig.\ref{fig:6}b). The structure
factor value $S_{MM}(0)$ rises  from  2.1 to 13 (Fig.\ref{fig:7}b).
For the case of pentavalent ions, the rdf peak rises from 150  to 230
(Fig.\ref{fig:6}c),   while the  initial  structure factor peak from 6
to 13 (Fig.\ref{fig:7}c). One can see from the Fig.\ref{fig:6}bc and
Fig.\ref{fig:7}bc  that the aggregation is much stronger expressed in
both 60:3 and 60:5 electrolytes than in the corresponding systems B
and C. This drastic change can be attributed to (i) the decrease of
the effective correlation parameter $\Gamma$ for the multivalent ions
in presence of  monovalent ones and (ii) the change in the macroion
effective charge, which has slightly higher magnitude for the systems
with monovalent ions present. The accumulation of these small but
finite macroion charges opposes the cluster-cluster aggregation and
prevents formation of very large clusters. 

\section{Conclusion}  
 
We  performed a  numerical study  of a  highly  asymmetric electrolyte
solution  containing strongly  charged macroions and different amounts
of salt. The addition of a simple monovalent salt gradually lowers the
effective macroion charge and reduces  the   solution  stability. 
Addition  of   small  amounts  of multivalent  salt produces  similar
effect.  The dose  of multivalent salt  that  exceeds  the  macroion 
isoelectric  concentration  causes macroion  charge  inversion   and 
electrostatically  driven  macroion aggregation.  At the isoelectric 
point, we observe formation of large macroion clusters. When the 
inverted macroion charge  becomes large enough,  the clusters
re-dissolve.   Enlarging the  counterion valency strenghens the
instability at the isoelectric point but stabilises the solution at 
the overdosed conditions. The  obtained stability diagram closely
resembles polyelectrolyte-induced colloidal aggregation, while the  
aggregation  mechanism  is   related  solely   to  electrostatic
correlation forces rather than the van der Waals attractions.

We are  grateful to Per  Linse, Boris Shklovskii, Toan  Nguyen, Michal
Borkovec,  Christian Holm,  and Peter  Schurtenberger  for stimulating
discussions. This work was financially supported by the Swiss National
Science Foundation.

\end{document}